\documentclass[superscriptaddress,prx,reprint,amsmath,twocolumn,amssymb]{revtex4-1}

\usepackage{graphicx} 
\usepackage{amsmath} 
\usepackage{xifthen}
\usepackage{physics}
\usepackage{dsfont}

\usepackage{printlen}
\usepackage{hyperref} 
\usepackage[noabbrev]{cleveref} 

\crefname{appendix}{appendix}{appendices}
\Crefname{appendix}{Appendix}{Appendices}
\usepackage{etoolbox} 
\makeatletter
\appto{\appendix}{%
  \@ifstar{\def\theequation@prefix{A.}}%
          {}%
}
\makeatother

\usepackage[load=prefixed,load=abbr,separate-uncertainty=true]{siunitx} 
\DeclareSIUnit{\pp}{\textup{p.p.}}

\newcommand{\Ham}{\hat{\mathcal{H}}}

\newcommand{\sub}[1]{_{\mathrm{#1}}}

\newcommand{\submath}[2]{_{\mathrm{#1}{#2}}}

\newcommand{\ii}{\mathrm{i}}
\newcommand{\ie}{i.e.\@}

\newcommand{\NbyN}[2]{{#1}${\times}${#2}}

\newcommand{\plus}{\texttt{+}}
\newcommand{\minus}{\texttt{-}}
\newcommand{\plusminusRaw}{%
	\ooalign{%
  		\raisebox{.1\height}{+}\cr%
  		\smash{\raisebox{-.6\height}{-}}\cr
 		}%
  		}%
\newcommand{\plusminus}{\texttt{\plusminusRaw}}

\newcommand{\etal}[2]{
	\ifthenelse{\equal{#1}{}}
		{\emph{et~al.}}
		{\emph{et~al}\cite{#1}.}
	}

\newlength{\subLen}
\AtBeginDocument{\settoheight{\subLen}{$x$}}

\newcommand{\X}{\textsc{X}}
\newcommand{\Y}{\textsc{Y}}
\renewcommand{\H}{\textsc{H}}
\newcommand{\V}{\textsc{V}}

\newcommand{\dash}{^{\prime}}

\newcommand{\hatd}[1]{\hat{#1}^{\dag}}
\newcommand{\creop}{\hatd{a}}
\newcommand{\annop}{\hat{a}}

\begin{document}

\title{Pushing Purcell enhancement beyond its limits}

\author{Thomas D.\ Barrett}
\email{thomas.barrett@physics.ox.ac.uk}
\author{Thomas H.\ Doherty}
\author{Axel Kuhn}
\email{axel.kuhn@physics.ox.ac.uk}
\affiliation{University of Oxford, Clarendon Laboratory, Parks Road, Oxford  OX1 3PU, UK}

\date{\today}

\begin{abstract}

Purcell-enhanced photon emission into a cavity is at the heart of many schemes for interfacing quantum states of light and matter. We show that the intra-cavity coupling of orthogonal polarisation modes in a birefringent cavity allows for the emitter and photon to be decoupled prior to emission from the cavity mode, enabling photon extraction efficiencies that exceed the, previously considered fundamental, limits of Purcell enhancement. Tailored cavity birefringence is seen to mitigate the tradeoff between stronger emitter-cavity coupling and efficient photon extraction, providing significant advantages over single-mode cavities. We then generalise this approach to show that engineered coupling between states of the emitter can equivalently `hide' the emitter from the photon, ultimately allowing the extraction efficiency to approach its fundamental upper limit. The principles proposed in this work can be applied in multiple ways to any emitter-cavity system, paving the way to surpassing the traditional limitations with technologies that exist today.
\end{abstract}

\maketitle

\section{Introduction}

Qubits encoding either a single unit of quantum information or part of a larger quantum state are fundamental building blocks of quantum technologies.  Indeed, many platforms primarily differ by the species of qubit chosen, from naturally occurring single atoms~\cite{saffman16, reiserer15, boozer98} and ions~\cite{cirac95, schafer18, wright19} to fabricated quantum dots~\cite{veldhorst14, yoneda18, watson18} and superconducting qubits~\cite{wallraff04, gambetta17, scarlino19}.  However, despite the significant progress made using matter-based qubits, when considering distributed quantum applications optical technologies remain the most promising platform due to the high speed and long coherence time of photonic states.  An efficient interface between quantum states of light and matter is then essential to the development of scalable quantum networks.

Optical cavities can provide this interface as they greatly enhance the interaction of any emitter placed within them to the resonant electric field modes.  Coupled emitter-cavity systems are then a versatile quantum technology with applications ranging from single photon production~\cite{press07, bochmann08, kang11, aharonovich16, dolan18, kuhn02, mckeever04}, over the photonic readout of a matter-based qubit state~\cite{stute13, keller17}, to entanglement distribution and networked quantum information processing (QIP)~\cite{kimble08, cirac98, fruchtman16, reiserer15}.  In this work we consider the most fundamental manifestation of emitter-cavity coupling, Purcell-enhanced spontaneous emission, and ultimately show that the widely accepted trade-offs associated with photon extraction efficiency and cavity design can be mitigated.

 Einstein's  phenomenologically derived discussion of spontaneous emission describes a behaviour so fundamental it is often considered an immutable property of matter~\cite{einstein17}.  However, a quantum mechanical treatment of this interaction, including the action of a vacuum field, shows that spontaneous emission is affected by the environment of the emitter.   In this light, Purcell first proposed that spontaneous emission could be enhanced within an electromagnetic resonator~\cite{purcell46} and laid the foundations for the field of cavity quantum electrodynamics (CQED).  The spontaneous emission rate of a quantised emitter, which we will herein interchangeably refer to as `atoms', in a cavity is enhanced by the Purcell factor~\cite{purcell46},
\begin{equation}
	f\sub{P} = \frac{3 \lambda^{3}}{4\pi^{2}}\frac{Q}{V},
	\label{eq:PurcellFactor}
\end{equation}
where $V$ is the resonator mode volume, $Q$ its quality factor and $\lambda$ the emission wavelength.  In CQED, this is given by $f\sub{P}=2C=g^{2}/(\kappa\gamma)$, where $g$ is the atom-cavity coupling rate, $\kappa$ is the cavity field decay rate, $2\gamma$ is the spontaneous emission rate of the uncoupled atom and $C$ is the cooperativity.  The fraction of emission which goes into the cavity mode is given by $f\sub{P}/(f\sub{P}+1)$.  Increasing emission into the cavity then requires working in the strong-coupling regime~\cite{kimble98}, typically defined as $C>1$, where the rate of excitation exchange between the emitter and the cavity dominates the loss of excitation from the system.  However, to efficiently map this emission into the cavity to useful photon emission from the cavity into a well-defined mode, the cavity decay rate, $\kappa$, must dominate the atomic decay, $\gamma$.  Whilst these conditions are not mutually exclusive, it is challenging to realise a system with both $C\gg1$ and $\kappa\gg\gamma$, especially within the constraints of feasible cavity designs~\cite{hunger10,gallego16} and mirror coatings within the optical regime~\cite{rempe92,hood01}.  More generally, this defines a trade-off between realising strong light-matter interactions (\ie{} where the interaction in the cavity is sufficiently sustained to allow coherent manipulation of the system) and efficient photon extraction.

In contrast to fast-excitation schemes where a strong pump pulse rapidly excites the emitter to stimulate photon emission~\cite{bochmann08,kang11}, single photon sources based on Raman transitions partly driven by the cavity field attempt to circumvent this issue by only transiently populating the excited atomic state~\cite{kuhn02, mckeever04}.  This allows efficient photon extraction even with $\kappa\leq\gamma$, however these adiabatic processes are slow and therefore not suitable for tasks such as enhanced fluorescence collection or high bandwidth single photon sources.  An alternative approach is to pursue stronger-coupling cavities that allow for faster photonic decay whilst maintaining $\kappa < g$.  This has motivated the development of microcavity resonators with tightly confined optical modes, such as Fabry-Perot cavities formed between the laser-ablated tips of optical fibres~\cite{steinmetz06,hunger10,gallego16,ruelle18}.  Those micron-scale mirrors are especially prone to birefringence -- a lifting of the degeneracy of their polarisation eigenmodes -- arising from any elliptical curvature of the mirror surface~\cite{uphoff15}.  In our recent work~\cite{barrett19} the effect of birefringence has been observed through the emission of single photons with time-dependent polarisation states.  This behaviour is explained by a simple model of direct coupling between orthogonal polarisation modes within the cavity.

Many previous studies have considered the limits of CQED~\cite{bozhevolnyi16, goto18} and demonstrated the coupling of various emitters -- including atoms~\cite{gallego18}, ions~\cite{steiner13, takahashi18}, molecules~\cite{toninelli10}, NV-centres~\cite{albrecht13, kaupp16, riedel17}, quantum dots~\cite{muller09,sanchez13,snijders18,liu18} and carbon nanotubes~\cite{hummer16,jeantet16} -- to microcavities.  However such work has considered single cavity modes to couple only to well-defined transitions of the emitter and to the extracted photon.  Therefore, the introduction of all-optical intra-cavity dynamics in CQED represents a paradigm shift.  The experimentally validated model of cavity birefringence allows us to challenge the standard limits of coherent photon extraction, as governed by the interplay of $g$, $\kappa$ and $\gamma$, via the introduction of an active manifold of coupled modes.

In this work we show that the presence of birefringence does not necessarily impair photon extraction from emitter-cavity systems.  In fact, in suitable regimes it can provide significant improvements.  These effects are shown to have the potential to impact systems with realistic parameters found in present-day experiments.  Also, we describe how the reduced coupling strength of longer confocal cavities -- a desirable regime in systems where dielectric mirrors interfere with the trapping-potential of an emitter between them -- can be accounted for with suitable birefringence.   This allows for equivalent performance -- compared to shorter-length, stronger-coupling counterparts -- to be achieved.  Furthermore, we discuss other approaches, inspired by our understanding of the mechanism behind the birefringence-enhanced effects, that modify either the atomic structure or intra-atomic couplings to achieve enhanced photon extraction efficiencies.  These are shown to provide equivalent or even better performance to birefringence-only enhancement, with the tradeoffs between these approaches discussed, and in principle allow the efficiency to approach the fundamental upper bound.
\section{Two-level atom}

\subsection{Two-level atom in a cavity}

We begin with the simplest example of a two-level atom coupled to a non-birefringent cavity, as a reference point to the novel systems we will then consider. The system is described by the Hamiltonian
\begin{equation}
\begin{aligned}
	\Ham/\hbar = &\Delta\sub{C} \ket{e}\bra{e} - g ( \ket{e}\bra{u} \annop + \creop\ket{u}\bra{e} ) \\
	&- \ii (\gamma \ket{e}\bra{e} + \kappa \creop\annop),
	\label{eq:hamTwoLevel}
\end{aligned}
\end{equation}
where $\ket{u}$ and $\ket{e}$ are the ground and excited atomic states, respectively, and $\Delta\sub{C}$ is the detuning of the cavity from atomic resonance.  In this model we consider emission to occur via either decay of the atomic amplitude at rate $\gamma$ or the decay of the cavity field at rate $\kappa$, which correspond to spontaneous emission from the atom into free space and cavity emission into a well defined mode, respectively.

\begin{figure}[t]
\includegraphics{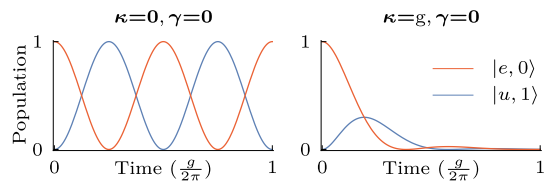}
\caption{Time evolution of a two-level atom coupled to a cavity.  The Rabi oscillations of the coupled system are shown for both the lossless ($\kappa{=}0$, $\gamma{=}0$) and critically damped ($\kappa{=}g$) cases.}
\label{fig:TwoLevelAtomEmission}
\end{figure}

With the atom initialised into the excited state, the photon extraction efficiency is readily given by the total probability of finding the system being in the ground state multiplied by the single-photon emission rate ($2\kappa$).  Optimum extraction requires a resonant cavity ($\Delta\sub{C} = 0$), in which case this extraction efficiency becomes,
\begin{equation}
	\eta\sub{ext} = \frac{\kappa g^2} {(\kappa + \gamma) (\kappa \gamma + g^2)} = \frac{\kappa}{\kappa + \gamma}\frac{2C}{2C + 1}.
\label{eq:eta}
\end{equation}
It is then straightforward to show that, for a given $g$ and $\gamma$, the maximum extraction efficiency, found when $\kappa=g$, is
\begin{equation}
	\eta\submath{ext,max} = \frac{g^2}{(g+\gamma)^{2}}.
	\label{eq:max}
\end{equation}

We emphasise that the Purcell-enhanced emission into the cavity mode, $f\sub{P}/(f\sub{P}+1)=2C/(2C+1)$, is only equivalent to the total cavity emission in the \emph{fast-cavity regime} where $\kappa \gg \gamma$.  It is significant that maximising the cooperativity of the atom-cavity system -- a measure of how efficiently and coherently the cavity can mediate light-matter interactions -- is not equivalent to maximising the efficiency with which we can extract photons (i.e. information) from the coupled system.  In the strong-coupling regime, with $C\gg1$, where sustained light-matter interactions can be realised, \cref{eq:eta} becomes
\begin{equation}
	\eta\submath{ext,}{C\gg1} = \frac{\kappa}{\kappa + \gamma}.
	\label{eq:etaStrongCoupling}
\end{equation}

Practically this tells us that there is not a single metric one can apply to tell whether a cavity is `better' or `worse' than another -- optimum cavity design is always subject to tradeoffs.  These are encapsulated in \cref{fig:TwoLevelAtomEmission} by considering the damped Rabi oscillations of the system as a photon is emitted.  Rapid transfer of the excitation from the atom to the cavity mode requires increasing the coupling strength, $g$.  However the cavity emission rate, $\kappa$, must trade-off rapid photon emission with sufficient build up of the cavity field to further stimulate the $\ket{e,0} \rightarrow \ket{u,1}$ transition.  Optimum photon extraction, stated already to be at $\kappa=g$, can then be understood as critical damping of the Rabi oscillations in the atom-cavity system when $\gamma=0$.

\subsection{Birefringence-enhanced photon extraction}
\label{sec:BirefringenceEnhancedPhotonExtraction}

\begin{figure}[t]
\includegraphics{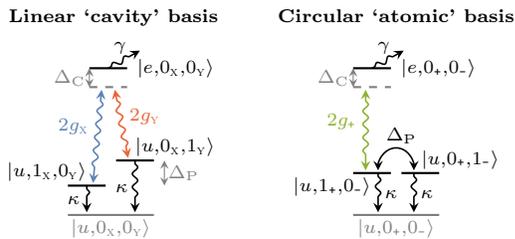}
\caption{A two level atom, with a circularly polarised transition, coupled to a birefringent cavity with linearly polarised eigenmodes.  In the cavity basis the circularly polarised photon emission couples to both cavity eigenmodes, which are split in energy by $\Delta\sub{P}$.  In the atomic basis, the photon is emitted into the correspondingly circular $\ket{\plus}$ cavity mode, which is coupled -- by the dephasing of its linearly polarised components -- to the orthogonal $\ket{\minus}$ mode at a rate $\Delta\sub{P}$.}
\label{fig:TwoLevelAtom}
\end{figure}

We now consider how birefringence modifies the photon emission properties of the atom-cavity system, and, in particular, how this additional degree of freedom allows us to outperform the limiting tradeoffs in cavity design previously discussed.  Birefringence is a lifting of the degeneracy of two orthogonal polarisation modes supported by the cavity.  As light circulates the cavity, these modes accumulate a relative phase difference at a rate $\Delta\sub{P}$, the frequency splitting between them. The result is that the polarisation state of light circulating the cavity rotates at a rate $\Delta\sub{P}$, unless it is aligned with one of the polarisation eigenmodes.  Recently this effect was directly observed through the emission of single photons from a birefringent cavity, with time-dependent polarisation states that evolved along their wavepacket~\cite{barrett19}.

For convenience, we take the polarisation eigenmodes of our cavity to be linearly polarised, $\{\ket{\X},\ket{\Y}\}=\{\ket{\H},\ket{\V}\}$, and consider the limiting cases of minimal and maximal overlap between this `cavity' basis and the `atomic' basis aligned with the coupled atomic transition from which a photon can be emitted into the cavity.  If the emitted photon is aligned with one of the cavity eigenpolarisations, the system reduces to the simple non-birefringent case already discussed.  However, in the case that the atomic basis is not aligned with the cavity polarisation basis, the effective coupling between orthogonal polarisation modes provides us with a mechanism to realise intra-cavity dynamics affecting only the photonic part of the atom-cavity system.

\begin{figure}[t]
\includegraphics{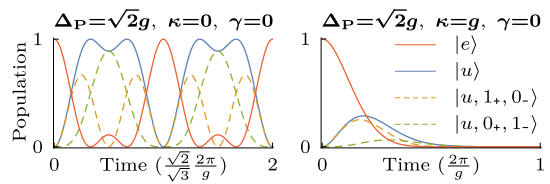}
\caption{Time evolution of a two-level atom coupled to a birefringent cavity.  The Rabi oscillations of the coupled system are shown for both lossless ($\kappa{=}0$, $\gamma{=}0$) and damped ($\kappa{=}g$) cases, with the birefringence, $\Delta\sub{P}=\sqrt{2}g$ chosen to maximise the time spend in the $\ket{u,\dots}$ states.}
\label{fig:TwoLevelAtomInBirefCavityEmission}
\end{figure}

\begin{figure*}[t]
\includegraphics{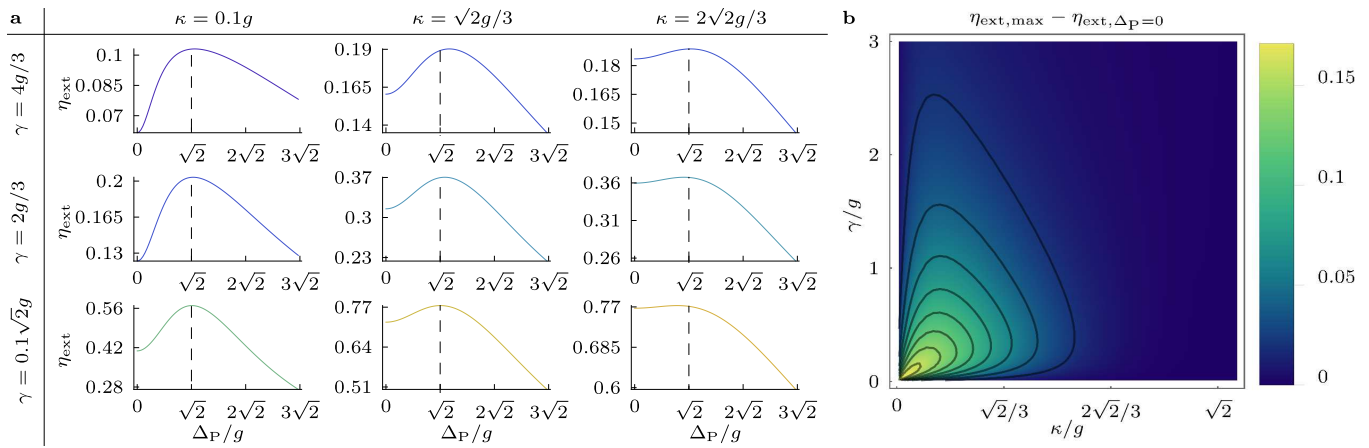}
\caption{Birefringence enhanced photon extraction from a two level atom.  (a) The photon extraction efficiency, $\eta\sub{ext}$, as a function of the cavity polarisation eigenmode splitting, $\Delta\sub{P}$, for a range of cavity field decay, $\kappa$,  and atomic amplitude decay, $\gamma$, rates.  The vertical dashed lines at $\Delta\sub{P}/g=\sqrt{2}$ denote the birefringence at which the coupled system is optimally biased towards populating photon-emitting states in the limit of strong coupling.  (b)  The emission enhancement achievable at given $\kappa/g$ and $\gamma/g$ by optimally tuning the birefringence, $\eta\sub{ext,max}$, in comparison to the non-birefringent case, $\eta\submath{ext,}{\Delta\sub{P}{=}0}$.  The contours are at multiples of 0.02.}
\label{fig:TwoLevelAtomBirefringenceEnhancedEmission}
\end{figure*}

Taking the quantisation axis of the system to be defined by the cavity axis, it is reasonable to consider the atomic transition to be coupled by circularly polarised light.  This is because, within the paraxial approximation, Fabry-Perot cavities can not support $\pi$-polarised light as this would require the electric field vector to be aligned with the direction of propagation.  The emission of a circularly polarised photon into a cavity with linearly polarised eigenmodes is summarised in  \cref{fig:TwoLevelAtom}, with the dressed states of the system expressed in either the circular basis, $\{\ket{\plus}=(\ket{\X}+i\ket{\Y})/\sqrt{2},\ket{\minus}=(\ket{\X}-i\ket{\Y})/\sqrt{2}\}$, or the linear basis.

Consider the case with a resonant cavity, $\Delta\sub{C}=0$ (which for a birefringent cavity we define as when each polarisation eigenmode is symmetrically detuned from atomic resonance by ${\pm}\Delta\sub{P}/2$).  The system dynamics can be seen in \cref{fig:TwoLevelAtomInBirefCavityEmission}, where an initial photon emission into $\ket{\plus}$ is coupled into the orthogonal $\ket{\minus}$ mode, where it is decoupled from the atom.  As a result, the atom spends longer in its ground state during these oscillations, only re-exciting as the photon polarisation continues to evolve back into $\ket{\plus}$.
The birefringence chosen, $\Delta\sub{P}=\sqrt{2}g$, maximises the amount of time the atom spends in the ground state (and correspondingly that a photon is in the cavity), which in this system corresponds to a $2/3$ of the overall time when averaged over many Rabi cycles.
By partially `hiding' the photon from the atom in an uncoupled mode, we allow the system to spend more time in states with the cavity mode populated, resulting in a higher upper bound for the photon extraction efficiency.  With a non-zero cavity emission rate, the oscillation between the orthogonal cavity modes would result in a corresponding oscillation in the polarisation of the emitted photon (the practical implications are discussed in \cref{sec:Discussion}).

To further examine birefringence-enhanced emission, \cref{fig:TwoLevelAtomBirefringenceEnhancedEmission}a shows the extraction efficiency as a function of the cavity polarisation eigenmode splitting for decay rates ranging from $\{\kappa,\gamma\} \ll g$ to $\{\kappa,\gamma\} \approx g$.  It is striking that for a given set of $\{g,\kappa,\gamma\}$, increasing only the cavity birefringence achieves enhanced photon extraction.  Moreover, this is a practical consideration as birefringence can be tailored independently of other rates in the system via the ellipticity of the mirrors~\cite{uphoff15,cui18,garcia18}.  The improvement in photon extraction is most significant in the strong-coupling regime.  This agrees with our intuitive understanding that for the additional coupling provided by birefringence to be impactful, the excitation initially stored in the atom must sufficiently populate the cavity mode and remain there for long enough that the intra-cavity coupling takes effect (\ie{} neither emission into free space, $\gamma$, nor from the cavity, $\kappa$, depopulates the system before intra-cavity coupling between polarisation modes can take effect).

To formalise this enhancement, recall that the photon extraction efficiency in the strong-coupling regime of a non-birefringent system (\cref{eq:etaStrongCoupling}) reduces to $\kappa/(\kappa + \gamma)$.  Birefringence allows a cavity excitation to circulate between two modes, one coupled to and the other uncoupled from the atom.  A suitable choice of $\Delta\sub{P}$ results in the atom-cavity system spending twice as long in states with a photon in the cavity than with an excited atom -- exactly the effect already seen in the undamped Rabi oscillations shown in \cref{fig:TwoLevelAtomInBirefCavityEmission}.  When considering only the overall extraction efficiency, doubling the time the system spends in a state from which it can emit a photon out of the cavity is equivalent to doubling the cavity decay rate.  We can therefore modify \cref{eq:etaStrongCoupling} to give the optimum extraction from our birefringent cavity,
\begin{equation}
	\eta\submath{ext,}{C\gg1,\Delta\sub{P}=\sqrt{2}g} = \frac{2\kappa}{2\kappa + \gamma},
	\label{eq:etaStrongCoupling2lvlBiref}
\end{equation}
where, as noted in the subscript, this optimum is found at $\Delta\sub{P}\approx\sqrt{2}g$ (tending to equality for increasing cooperativities).  The maximum achievable increase in extraction efficiency is then
\begin{equation}
	\eta\submath{ext,}{C\gg1,\Delta\sub{P}=\sqrt{2}g} - \eta\submath{ext,}{C\gg1,\Delta\sub{P}=0\vphantom{\sqrt{2}}} = \frac{\kappa\gamma}{(\kappa + \gamma) (2\kappa+\gamma)}.
\end{equation}
The upper limit on this enhancement is then an additional $3-2\sqrt{2} \approx \SI{17.2}{\pp}$ (\si{\pp} denotes percentage points) photon emission, found when $\kappa=\gamma/\sqrt{2}$.  These considerations can be seen in \cref{fig:TwoLevelAtomBirefringenceEnhancedEmission}b, which plots the (numerically found) maximum achievable enhancement, relative to the non-birefringent case, by tuning only the polarisation eigenmode splitting of the cavity, as a function of the decay rates of the system.

\begin{figure}[t]
\includegraphics{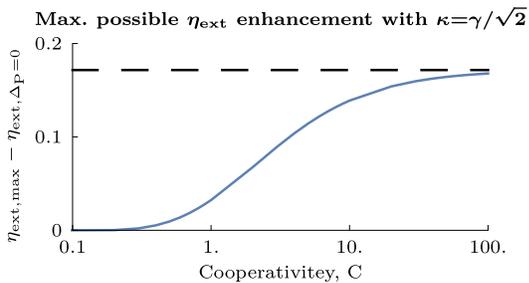}
\caption{Maximum achievable enhancement to photon extraction efficiency as a function of cooperativity, $C$, for a two-level atom coupled to a birefringent cavity.  Note the logarithmic scaling of the cooperativity axis.  For a given cooperativity, the maximum enhancement is possible when $\kappa=\gamma/\sqrt{2}$ and is achieved at $\Delta\sub{P}=\sqrt{2}g$.  For stronger coupling systems, the enhancement tends to an upper limit of ${\approx}\SI{17.2}{\pp}$, which is shown as the dashed line.}
\label{fig:TwoLevelAtomMaximumBirefringenceEnhancedEmissionVsCooperativity}
\end{figure}

Whilst the benefits of birefringence are clear in the strongly-coupled regime, it is of interest to investigate how the effect scales with cooperativity, $C$.  Considering a system with $\kappa=\gamma/\sqrt{2}$ and $\Delta\sub{P}=\sqrt{2}g$, \cref{fig:TwoLevelAtomMaximumBirefringenceEnhancedEmissionVsCooperativity} shows the increased photon extraction efficiency, in comparison to the non-birefringent case, as a function of $C$.  As expected we see no significant benefit from birefringence for weakly coupling systems, with very strongly coupling systems tending towards the upper limit of a \SI{17.2}{\pp} increase in efficiency.  Significantly however, even systems with cooperativities inline with those regularly demonstrated in the lab~\cite{kimble08} see non-negligible enhancement, with $C=\ $\numlist{1;5;10} corresponding to enhancements of \SIlist{3.2;11.4;13.9}{\pp}, respectively.

\section{$N$-level atom}



\begin{figure*}
\includegraphics{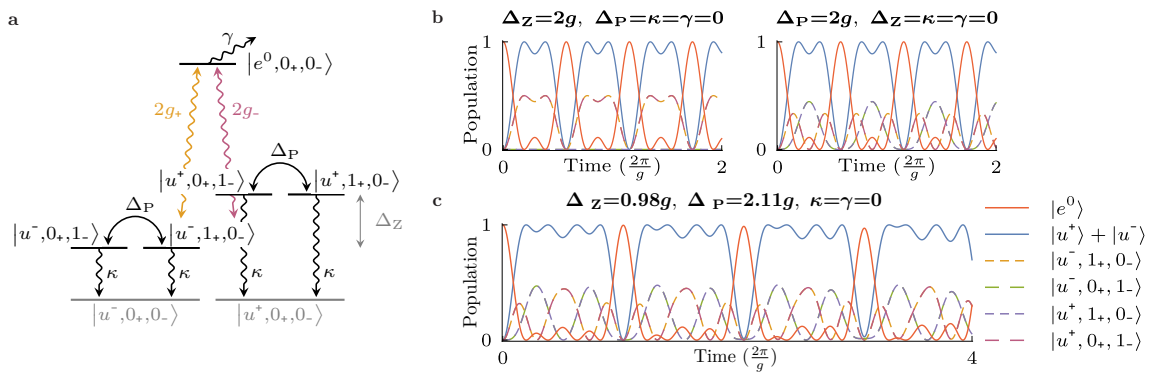}
\caption{Three-level atom coupled to a birefringent cavity.
(a) Circularly polarised atomic transitions, $\ket{u^{\plus}}\leftrightarrow\ket{e^0}$ and $\ket{u^{\minus}}\leftrightarrow\ket{e^0}$ coupled to a birefringent cavity with linearly polarised eigenmodes.  The ground states of the atom are split in energy by $\Delta\sub{Z}$ and the cavity polarisation eigenmodes are split in energy by $\Delta\sub{P}$, resulting in an equivalent coupling between $\ket{\plus}$ and $\ket{\minus}$ photon modes of the cavity.
(b) The Rabi oscillations of the coupled system considering only the effects of non-degenerate atomic ground states or cavity birefringence, with the ground state splitting, $\Delta\sub{Z}=2g$, or birefringence, $\Delta\sub{P}=2g$, respectively chosen for each case to maximise the time spend in $\ket{u^{\plus}}+\ket{u^{\minus}}$.
(c) The system's evolution with both the cavity birefringence and the atomic ground state splitting chosen to maximise the time spend in $\ket{u^{\plus}}+\ket{u^{\minus}}$ within the presented time interval.}
\label{fig:ThreeLevelAtom}
\end{figure*}

We have seen that cavity birefringence can be used to populate a state within an atom-cavity system where the atomic transition is decoupled from light circulating within the cavity mode.  Still, the bi-directional nature of the coupling to and from this state ultimately provides an upper limit to the achievable enhancement of cavity emission.  For example, a two-level atom coupled to a birefringent cavity has two dressed states in which the excitation populates the cavity mode, and so we are at best able to bias the system such that a photon in the cavity is decoupled from the atom \SI{50}{\percent} of the time.  It then stands to reason that if we coupled to more dressed states where the photon is decoupled from the cavity, we could further enhance the photon extraction efficiency.  Whilst birefringence alone cannot achieve this, in this section we consider the effects of multiple atomic ground levels with birefringence induced, and birefringence inspired, couplings between them.

\subsection{Three-level atom}

Consider extending the two-level atomic model considered so far to a three-level atom with two ground states, $\ket{u^{\plus}}$ and $\ket{u^{\minus}}$, and a single excited state $\ket{e^{0}}$, where the superscripts denote the magnetic spin of each level.  
If we once again consider a birefringent cavity with linear eigenpolarisations, then we have the system shown in \cref{fig:ThreeLevelAtom}a, where $\Delta\sub{Z}$ is the energy splitting of the atomic ground states (a convention chosen in reference to the Zeeman effect, where an external magnetic field lifts the degeneracy of such magnetic sublevels).  We emphasise at this point that this system is not overly contrived but can, in fact, arise in any atom-cavity coupling scheme where the ground state of the atom has a non-zero angular momentum quantum number~\cite{barrett18b, wilk07}.

Consider the system in two limiting cases; with only birefringence ($\Delta\sub{P} \neq 0$, $\Delta\sub{Z} = 0$) or detuning ($\Delta\sub{P} = 0$, $\Delta\sub{Z} \neq 0$) induced couplings.  Though these cases will differ in the time-evolution of individual dressed states, the behaviours of both are equivalent when considering only the two relevant states for photon extraction, $\sum \ket{u^\plusminus,\dots}$ (\ie{} states allowing for emission from the cavity) and $\ket{e^0,0,0}$ (see \cref{app:threeLevelAtomModel} for details).  This is manifest in \cref{fig:ThreeLevelAtom}b which shows the undamped Rabi oscillations for an energy splitting of $2g$, chosen to maximise the time-averaged population in the photon emitting states, in each case.

By further consideration of the case with no detuning of the atomic ground levels ($\Delta\sub{P} \neq 0$, $\Delta\sub{Z} = 0$), it can be seen that the couplings to and from states with a photon in the cavity are equivalent regardless of whether the atom is in $\ket{u^{\plus}}$ or $\ket{u^{\minus}}$.  States with the same photonic configuration (the unique cases being $\ket{u^\plusminus,1\sub{\plus},0\sub{\minus}}$ and $\ket{u^\plusminus,0\sub{\plus},1\sub{\minus}}$) then evolve identically in time, allowing the system in these limiting cases to be effectively reduced to a two-level atom in a birefringent cavity, as was considered in \cref{sec:BirefringenceEnhancedPhotonExtraction}.   A rigorous treatment shows that this parallelism holds up to a scaling of the atom-cavity coupling rate, with equivalent effects observed for $2g$ in the the three-level case and $\sqrt{2}g$ in the two-level case.   This can be seen by comparing \cref{fig:ThreeLevelAtom}b to the undamped Rabi oscillations of the equivalent two-level system in \cref{fig:TwoLevelAtomInBirefCavityEmission}.

How a suitable birefringence `hides' a photon in the cavity in a mode decoupled from the atom has been discussed in \cref{sec:BirefringenceEnhancedPhotonExtraction}.  A physical justification for observing the same effect in a three-level atom with $\{\Delta\sub{P} = 0,\,\Delta\sub{Z} \neq 0\}$ relates to the opposite detuning of each ground atomic state from resonance with the cavity.  These detunings result in time-dependent phase terms of opposite sign for light circulating in each circular polarisation mode and so these modes periodically de-phase and re-phase.  When the atom equally populates the two ground levels and is driven by fields of opposite phase, destructive interference between these two paths inhibits the re-excitation.  For a suitably chosen detuning, every second Rabi oscillation is inhibited, maximising the amount of time the system spends with the atom in a ground state and a photon in the cavity.

Combining both cavity birefringence and detuning ($\Delta\sub{P} \neq 0$, $\Delta\sub{Z} \neq 0$) allows us to increase the photon extraction efficiency from our three-level atom beyond the limits of what could be achieved with only a two-level emitter.  In this situation, the atom-cavity system is sufficiently complex that we numerically find values of $\Delta\sub{P}$ and $\Delta\sub{Z}$ that maximally bias the system's population towards states from which the cavity can emit a photon within a time interval of interest.  \Cref{fig:ThreeLevelAtom}c shows the complex oscillation dynamics within the system for $\Delta\sub{P}{=}2.11g$ and $\Delta\sub{Z}{=}0.98g$, values which maximise the cavity mode population within the plotted $8\pi/g$ time window.  


As we consider ever longer time intervals, the ratio of time the system spends in a state from which a photon can be emitted from the cavity ($\sum \ket{u^{\plusminus},\dots}$) to those where it cannot ($\ket{e^0,0\sub{\plus},0\sub{\minus}}$) tends to 4/5.  With four such photon-emitting dressed states in our system and a single non-emitting dressed state, this agrees with our previous argument that with symmetric couplings we can at best evenly distribute the time-averaged population across all coupled states.  Extending this to consider the efficiency with which we can extract an atomic excitation as a cavity photon, we assert that (analogously to case of a two-level atom in \cref{eq:etaStrongCoupling2lvlBiref}) in the limit of strong-coupling
\begin{equation}
	\eta\submath{ext,max,}{C\gg1} = \frac{4\kappa}{4\kappa + \gamma},
	\label{eq:etaStrongCoupling3lvlBirefDet}
\end{equation}
where the subscript `max' denotes $\Delta\sub{P}$ and $\Delta\sub{Z}$ chosen to maximise the extraction efficiency.  By comparison to the  simple case where $\Delta\sub{P}=\Delta\sub{Z}=0$ (described by \cref{eq:eta}) we find that the upper limit on this enhancement is ${\approx} \SI{33.3}{\pp}$, achievable when $\kappa=\gamma/2$, an even greater improvement than is provided by birefringence alone.
Although maximum enhancement is achieved in strong coupling, further investigation (see \cref{app:threeLevelSimulations}) shows that significant improvements are obtained even for realistic cooperativities that can be obtained in practice.  For $C{=}1$ up to ${\approx}$\SI{10}{\pp} extraction enhancement can be obtained, rising to ${\approx}$\SI{28}{\pp} for $C{=}10$.  This represents a significant improvement in readily obtained experimental regimes. 

\subsection{$n$ coupled ground states}

\begin{figure}[b]
\includegraphics{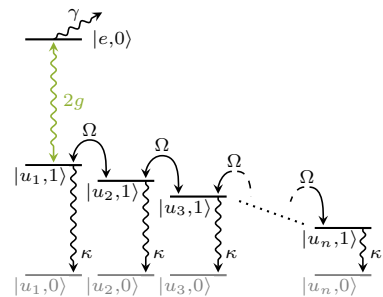}
\caption{An $N$-level atom, with a single transition coupled by a cavity at a rate $g$.  The $n{=}N{-}1$ ground states are coupled in a chain at a rate $\Omega$.  Each of the resulting dressed states, $\ket{u_i,1}$, decays to $\ket{u_i,0}$ at a rate $\kappa$ as the photon is emitted from the cavity.}
\label{fig:NLevelAtomModel}
\end{figure}

\begin{figure*}[]
\includegraphics{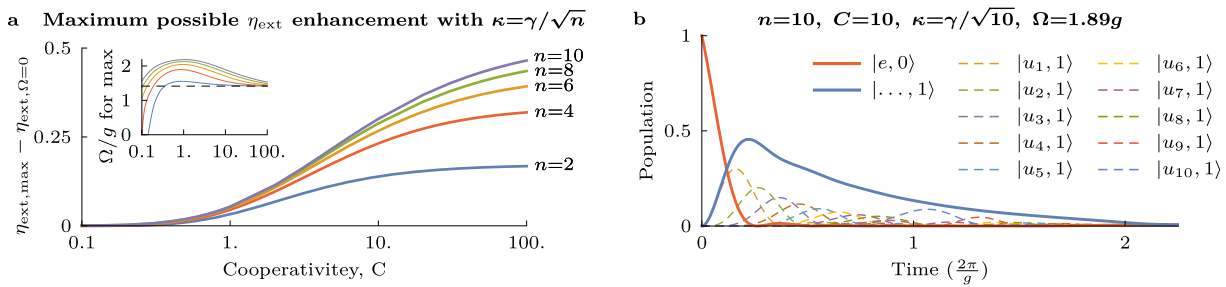}
\caption{Photon extraction from an $N$-level atom, with $n{=}N{-}1$ ground levels coupled to form a chain at a rate $\Omega$. (a) Maximum achievable photon extraction enhancement as a function of cooperativity, $C$.  Inset are the intra-atomic level coupling rates, $\Omega$, required for maximum enhancement.  In all cases this tends to $\Omega/g=\sqrt{2}$ (dashed line) for increasing cooperativities.  Note the logarithmic scaling of the cooperativity axes.  (b) The time evolution of the system with the optimal parameters for photon extraction at $n=10$, $C=10$.  The solid traces represent the single-state population in which the atom is excited (red) and the population in the other one-photon state traced over all $n$ remaining states (blue).  The individual population of these photon-emitting states are shown as the dashed traces.}
\label{fig:NLevelAtomEmissionEnhancement}
\end{figure*}

The effects we have discussed so far arise from additional couplings beyond those between the atomic states that are coupled by the cavity.  Birefringence provides an intra-cavity coupling mechanism between orthogonal polarisation modes, and detuning the ground levels of a three-level atom provides an equivalent coupling between the in-phase and out-of-phase superpositions of the photon-emitting dressed states.  In principle, increasing the number of coupled  states from which a cavity photon can be emitted increases the efficiency with which this emission can be realised -- however, with only two orthogonal polarisation modes supported within a cavity, it is difficult to continue to increase the number of coupled states by acting on the photonic part of the system alone.  By contrast, the emitter can have many energy states decoupled from the cavity and in this section we consider how equivalent couplings between these modifies emitter-cavity systems.

\Cref{fig:NLevelAtomModel} shows a general $N$-level atom, emitting into a non-birefringent cavity.  We assume the $n{=}N{-}1$ ground levels of the atom are coupled at a rate $\Omega$ to form a chain.  Such a system could be physically realised in various ways, for example by directly coupling magnetic sublevels or hyperfine levels of an atom with an RF field.  Alternatively, the precession of the atomic dipole moment around a magnetic bias field that is orthogonal to the cavity axis would directly couple the magnetic sublevels of the atom when viewed, as is the case for photon emission, in the cavity basis~\cite{wilk08}.  

Arguing once again that we can at best bias the system to equally populate all coupled levels, the maximum photon extraction efficiency in the strong coupling regime is then given by
\begin{equation}
	\eta\submath{ext,max,}{C\gg1} = \frac{n\kappa}{n\kappa + \gamma}.
	\label{eq:etaStrongCouplingMaxNlevel}
\end{equation}
In comparison to a two-level atom (with $n=1$, or equivalently \cref{eq:etaStrongCoupling}) the maximum possible enhancement is $1-2/(1+\sqrt{n})$, which is achieved when $\kappa=\gamma/\sqrt{n}$.  With two coupled ground levels, $n=2$, this is exactly equivalent to the two-level atom in a birefringent cavity discussed in \cref{sec:BirefringenceEnhancedPhotonExtraction}.

\Cref{fig:NLevelAtomEmissionEnhancement}a shows the maximum achievable enhancement of photon extraction as a function of cooperativity for $n=\{2,4,6,8,10\}$.  The impact of adding more coupled ground levels can be seen to diminish for larger $n$, and although these larger systems can provide more efficient extraction, this improvement requires increasingly high cooperativities to be significant.  Practically this reinforces that, as we have seen, relatively few coupled levels and physically realistic cooperativities are sufficient to realise significant enhancement.  Inset to the figure are the intra-level coupling rates, $\Omega$, that maximise the emission efficiency.  In all cases we see that these tend to $\Omega=\sqrt{2}g$ in the limit of increasing cooperativity.

The emission dynamics of the maximally enhanced $n=10$ system at $C=10$ are shown in \cref{fig:NLevelAtomEmissionEnhancement}b.  We can see that once the atomic excitation has transferred into the cavity, the atomic population cascades through the ground states, decoupled from re-excitation, greatly increasing the probability of a photon emission from the cavity mode.
\section{Discussion}
\label{sec:Discussion}

In this work we have discussed how cavity birefringence, $\Delta\sub{P}$, modifies the coupling dynamics of emitter-cavity systems beyond the hitherto anticipated limitations.  Accordingly, we present an approach to systematically improve the photon emission probability from a coupled atom-cavity system beyond what could be achieved using even a perfectly optimised non-birefringent system.  In birefringence-modified systems, the intra-cavity coupling of orthogonal polarisation modes allows photons to be `hidden' in a cavity mode decoupled from the emitter.  Far from inhibiting the efficiency of cavity-based light-matter interfaces, suitably applied birefringence is found to provide significant improvements. This has been shown in the most fundamental application of such an interface, the extraction of an atomic excitation into a useable emission from the cavity.

The prescience of this work is underlined by recent experimental work.  We showed that the effects of birefringence can become significant in the strong-coupling regime ($C > 1$) which has been recently demonstrated with both atoms~\cite{gallego18} and ions~\cite{takahashi18} in open microscopic cavities.  The ability to tailor non-negligible birefringence into these cavities has also recently been developed~\cite{garcia18,cui18}.  Therefore, the experimental regimes discussed within this work are within current technological limits and have the potential to significantly impact present-day and future experiments.

In these practical considerations, birefringence can be viewed as another free-parameter in cavity design, with its ability to mediate the tradeoff between sustained intra-cavity interactions ($g\gg\{\kappa,\gamma\}$) and extraction efficiency (sufficiently large $\kappa$).  As a concrete example, this allows for the construction of longer cavities without sacrificing performance.  A detailed discussion of this point is provided in \cref{app:cavityDesign}, where we conclude by analysing the impact of birefringence-enhanced photon extraction on a state-of-the-art ion-cavity system of Takahashi\etal{}~\cite{takahashi18}.  Even when considering the simplest implementation -- where we do not utilise any additional couplings between atomic states -- we find that tailored birefringence allows the cavity to be lengthened by over \SI{50}{\percent} without a loss of performance.  This is highly desirable when trapping ions as the interaction between the trapping fields and the dielectric mirrors of the cavity is a major experimental obstacle.  With such systems being leading candidates to form stationary matter-based processing nodes in quantum networks~\cite{fruchtman16, kimble08}, such an improvement in the viable parameter space is of immediate and significant importance.

Further approaches for increasing the number of states in which a photon in the cavity is decoupled from the emitter have also been discussed.  Specifically we considered a three-level emitter coupled in a $\Lambda$-system with ground states oppositely detuned from resonance with the cavity.  The polarisation modes that couple the atomic transitions accumulate a phase difference as a result of the detunings, periodically de-phasing such that destructive interference between the two possible transition pathways suppresses re-excitation of the emitter.  A more general model considering an emitter with multiple directly coupled ground states is then in essence inverting the effect of birefringence by `hiding' the emitter from a photon in the cavity mode.  In all cases, increasing the number of `hidden' states which the system can populate without coupling back to the excited atomic state provides an increased photon extraction probability.  Significantly, however, despite the increasing scale and complexity of these approaches, the majority of the benefits are realised in simple systems at physically realistic parameters.

From an implementation perspective, there are trade-offs to consider when increasing photon extraction efficiency by increasing the number of coupled states in either the photonic or atomic part of the emitter-cavity system.  The additional coupling distributes the population across multiple states, resulting in either the emission of photons with time-dependent polarisation states (\ie{} `impure' photonic qubits) or leaving the emitter distributed across multiple ground states (\ie{} `impure' atomic qubit).  These time-dependent polarisation states have already been measured in experiments to agree with a simple theoretical extension of the Jaynes---Cummings model~\cite{barrett19} and in principle could be corrected for after emission using, for example, a Pockels cell.  It should be noted, however, that modified polarisation states may not be problematic in many systems.  In the trivial case when polarisation control is not required, it is clear that the modified emission will not inhibit performance.  However, even if the emitted photons must be indistinguishable and cannot be `corrected' using a Pockels cell, in principle two highly birefringent systems can still produce perfectly indistinguishable photons as the emission dynamics are deterministic (and well understood).  Whilst it is typical to address this by minimising birefringence, our proposal simply suggest the systems are both tailored to the same non-birefringent configuration in order to reap additional benefits.

Finally, we emphasise that although the presented discussion considers an atom initialised in an excited state, the effects demonstrated are practical and the and intuition gained is readily extend to other physical systems.  Firstly, many systems excite the atom on timescales far faster than cavity effects (for example~\cite{bochmann08}), and are therefore well approximated by our models.  If the photonic emission is the result of an adiabatic population transfer, as is the case for V-STIRAP~\cite{kuhn02, mckeever04} single-photon emission, there remains an effective coupling between a state with a photon in the cavity and a state without a photon.  As such, by `hiding' either the cavity photon or the atom in an uncoupled state, using the techniques we present, it is still possible to realise enhanced emission.

Overall, we have demonstrated that the traditional limits to photon extraction efficiency can be surpassed by decoupling the emitter and the photon prior to emission from the cavity.  The passive nature of birefringence, it need only be engineered into the system during construction, and the fact that this simple system can provide substantial performance benefits, make it a promising tool for realising this enhancement.  However, the concept can be applied to any emitter-cavity system and, as we have shown, can be realised in multiple ways.  As such, these principles have the potential to impact any cavity-based platform and the multitude of applications these support.

\begin{acknowledgements}
The authors would like to thank Juan-Rafael \'Alvarez and M.\ IJspeert for their valuable input, and to acknowledge support for this work through the UK National Quantum Technologies Programme (NQIT hub, EP/M013243/1) and the EU-ITN LIMQUET.
\end{acknowledgements}

\newpage

\appendix
\crefalias{section}{appendix}
\section{Birefringent $\Lambda$-system}
\label[appendix]{app:threeLevelAtomModel}

The interaction Hamiltonian for the system shown in \cref{fig:ThreeLevelAtom}a is
\begin{equation}
	\frac{\Ham\sub{int}}{\hbar} {=} \pmqty{
		-\tfrac{\Delta\sub{Z}}{2}{-}\ii\kappa & -\tfrac{\Delta\sub{P}}{2} & 0 & 0 & -g \\
		-\tfrac{\Delta\sub{P}}{2} & -\tfrac{\Delta\sub{Z}}{2}{-}\ii\kappa & 0 & 0 & 0 \\
		0 & 0 & \tfrac{\Delta\sub{Z}}{2}{-}\ii\kappa & -\tfrac{\Delta\sub{P}}{2} & 0 \\
		0 & 0 & -\tfrac{\Delta\sub{P}}{2} & \tfrac{\Delta\sub{Z}}{2}{-}\ii\kappa & -g \\
		-g & 0 & 0 & -g & -\ii\gamma
		}.
\end{equation}
The change of basis to diagonalise the upper left \NbyN{4}{4} sub-matrix, corresponding to states with a photon in the cavity mode, is
\begin{equation}
\begin{aligned}
	& \ket{u^\plusminus} \creop\sub{\plus} \rightarrow \ket{u^{\texttt{\plusminus}}} (\creop\sub{\plus} - \creop\sub{\minus})/\sqrt{2}, \\
	& \ket{u^{\texttt{\plusminus}}} \creop\sub{\minus} \rightarrow \ket{u^{\texttt{\plusminus}}} (\creop\sub{\plus} + \creop\sub{\minus})/\sqrt{2}, \\
	& \ket{e^0} \creop\sub{\texttt{\plusminus}} \rightarrow \ket{e^0} \creop\sub{\texttt{\plusminus}}.
\end{aligned}
\end{equation}
This is given by the unitary transformation,
\begin{equation}
	U = \pmqty{ 
		1/\sqrt{2} & 1/\sqrt{2} & 0 & 0 & 0 \\
		-1/\sqrt{2} &  1/\sqrt{2} & 0 & 0 & 0 \\
		0 & 0 & 1/\sqrt{2} & 1/\sqrt{2} & 0 \\
		0 & 0 & -1/\sqrt{2} &  1/\sqrt{2} & 0 \\
		0 & 0 & 0 & 0 & 1
		},
\end{equation}
from which we find $\Ham\sub{int}\dash=U.\Ham\sub{int}.U^{\intercal}$,
\begin{widetext}
\begin{equation}
\frac{\Ham\sub{int}\dash}{\hbar} =	\pmqty{
		\tfrac{\Delta\sub{P}}{2}-\tfrac{\Delta\sub{Z}}{2}-\ii\kappa & 0 & 0 & 0 & -\tfrac{g}{\sqrt{2}} \\
		0 & \tfrac{\Delta\sub{P}}{2}-\tfrac{\Delta\sub{Z}}{2}-\ii\kappa & 0 & 0 & \tfrac{g}{\sqrt{2}} \\
		0 & 0 & -\tfrac{\Delta\sub{P}}{2}+\tfrac{\Delta\sub{Z}}{2}-\ii\kappa & 0 & -\tfrac{g}{\sqrt{2}}\\
		0 & 0 & 0 & \tfrac{\Delta\sub{P}}{2}+\tfrac{\Delta\sub{Z}}{2}-\ii\kappa & -\tfrac{g}{\sqrt{2}} \\
		-\tfrac{g}{\sqrt{2}} & \tfrac{g}{\sqrt{2}} & -\tfrac{g}{\sqrt{2}} & -\tfrac{g}{\sqrt{2}} & -\ii\gamma
		}.
	\label{eq:HintRot}
\end{equation}
\end{widetext}
It is clear from \cref{eq:HintRot} that the roles of $\Delta\sub{Z}$ and $\Delta\sub{P}$ are interchangeable when considering only the total population of cavity-emitting ($\sum \ket{u^\plusminus,\dots}$) and non-cavity-emitting ($\ket{e^0,0,0}$) states.  Moreover we see that the cavity coupling strengths are effectively modified by a factor of $1/\sqrt{2}$ in comparison to the case of a simple two-level atom coupled to a resonator.

\section{Photon extraction from a three-level atom in a birefringent cavity}
\label[appendix]{app:threeLevelSimulations}

\begin{figure}[!b]
\includegraphics{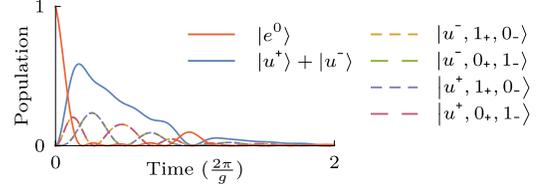}
\caption{Photon extraction from a three-level atom, with ground levels split by $\Delta\sub{Z}$, emitting into the $\ket{\plus}$ and $\ket{\minus}$ modes of a cavity with linearly polarised eigenmodes split by $\Delta\sub{P}$. (a) Maximum achievable photon extraction enhancement as a function of cooperativity, $C$.  Note the logarithmic scaling of the cooperativity axis.  For a given cooperativity, the maximum enhancement is possible when $\kappa=\gamma/2$.  For stronger coupling systems, the enhancement tends to an upper limit of ${\approx}\SI{33.3}{\pp}$, which is shown as the dashed line. (b) The time evolution of the system with the optimal parameters for photon extraction at $C=10$.}
\label{fig:ThreeLevelAtomEmissionEnhancement}
\end{figure}

In the main text we discuss enhanced photon extraction from a three-level atom coupled to a birefringent cavity (see \cref{fig:ThreeLevelAtom}) in the limit of strong-coupling ($C\gg1$).  \Cref{fig:ThreeLevelAtomEmissionEnhancement}a shows the maximum possible enhancement for systems with $\kappa=\gamma/2$ as a function of the cooperativity where we see that significant improvements are obtained even for even model cooperativities ($C{\sim}1$).  The emission dynamics of the atom-cavity system for the $C{=}10$ case with optimally chosen $\Delta\sub{P}$ and $\Delta\sub{Z}$ are shown in \cref{fig:ThreeLevelAtomEmissionEnhancement}b, where we see the suppression of atomic re-excitation once a photon has been emitted into the cavity.

\section{Implications for cavity design}
\label[appendix]{app:cavityDesign}

\begin{figure*}
\includegraphics{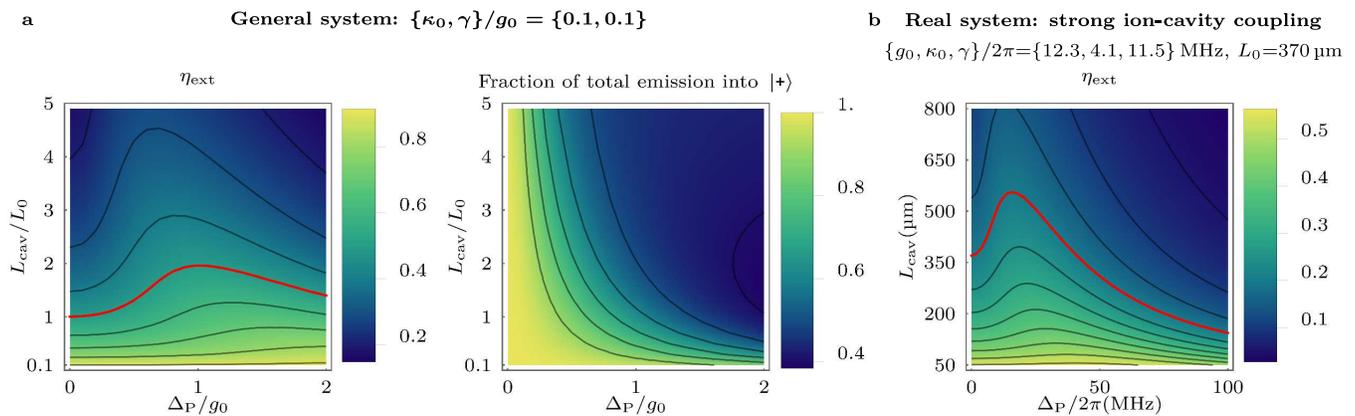}
\caption{Photon extraction properties for a two-level atom emitting circularly polarised, $\ket{\plus}$, photons into a cavity with linearly polarised eigenmodes split by $\Delta\sub{P}$.  (a) The photon extraction efficiency, $\eta\sub{ext}$, and polarisation purity for a strong-coupling system.  The contours of both plots are at multiples of 0.2. (b) An equivalent analysis of photon extraction efficiency considering the cavity implemented by Takahashi \emph{et al.}~\cite{takahashi18} to provide a physically realistic parameter set.  The contours are again at multiples of 0.2, with thick red contour at 0.2 denoting the efficiency realised by the chosen parameter set without birefringence as described in the text.}
\label{fig:TwoLevelAtomExtractionVsLength}
\end{figure*}

Control over the interaction of quantum states of light and matter realised with cavity resonators can be utilised in many systems.  However, the optimum cavity design is highly dependent on the desired application, as seen in the already discussed tradeoff between stronger light-matter interactions (high $g$, minimal $\kappa$) and fast, efficient photon extraction (higher $\kappa$).  Moreover, any realistic system will be subject to physical constraints, for example in the engineering limits of the mirror geometries and their dielectric coatings.  In this section we emphasise that engineered birefringence is not simply a mechanism for increased photon extraction, but more broadly should be considered an additional degree of freedom with which to improve cavity implementations and offset other manufacturing limitations. 

As an example, we consider the design of longer cavities, which is a desirable regime with emitters held in electro-magnetic fields, such as trapped ions or atoms. Long cavities are necessary to minimise the interference of the dielectric mirror surfaces with the trapping fields~\cite{keller16}. However, any increased cavity length, $L\sub{cav}$, comes with a reduced ion-cavity coupling, $g \propto L\sub{cav}^{-1/2}$, due to the increased cavity mode volume, and a reduced cavity emission rate, $\kappa \propto L\sub{cav}^{-1}$.  The rate of spontaneous emission into free space of an excited state is unchanged so the overall result of lengthening the cavity is to decrease the photon extraction efficiency.  Correspondingly, shortening the cavity in principle allows for more efficient photon extraction.  Formally we can define
\begin{equation}
	g = \frac{g_0}{\sqrt{L\sub{cav}/L_0}}, \qquad
	\kappa = \frac{\kappa_0}{L\sub{cav}/L_0},
	\label{eq:ratesVsLength}
\end{equation}
where $\g_0$ and $\kappa_0$ are the coupling rate and decay rate at some reference cavity length, $L_0$.  From this we can see that the cooperativity, $C=g^2/(2\kappa\gamma)$, is independent of cavity length.  However, the photon extraction efficiency decreases with increasing cavity length.  Note that any change to the cavity waist resulting from a change of cavity length can be corrected by correspondingly changing the radii of curvature of the mirrors.  For simplicity this effect is neglected in this discussion. 

We now consider a two level emitter of photons into the $\ket{+}$ polarisation mode, coupled to a birefringent cavity with linearly polarised eigenmodes (\ie{} the same system discussed in \cref{sec:BirefringenceEnhancedPhotonExtraction} and presented in \cref{fig:TwoLevelAtom}).  Starting in the strong coupling regime with $\kappa_0/g_0 = \gamma/g_0 = 0.1$, \cref{fig:TwoLevelAtomExtractionVsLength}a shows how the overall photon extraction efficiency, $\eta\sub{ext}$, changes as a function of cavity birefringence and length.  As expected, we see that for a given birefringence, $\Delta\sub{P}$, as the cavity is lengthened (shortened), the extraction efficiency decreases (increases).

Significantly however, the contours of constant $\eta\sub{ext}$ show that the cavity length and extraction efficiency can be partially decoupled by a suitable cavity birefringence, allowing equivalent extraction efficiencies to be realised at longer cavity lengths.  This indicates that the reduced $\kappa$ and $g$ are more than compensated for by biasing the systems Rabi oscillations, using birefringence, to spend relatively more time in states from which a cavity-photon can be emitted.  However, for longer cavities the coupling eventually becomes too weak to be fully compensated for in this manner, placing an upper limit on how long the cavity can be made without sacrificing performance.  For example, the thick red contour denotes $\eta\sub{ext}=0.5$ (the efficiency realised without birefringence from our considered system with $\kappa_0/g_0 = \gamma/g_0 = 0.1$ and $\Delta\sub{P}=0$ from \cref{eq:etaStrongCoupling}) and shows that we can at best approximately double the cavity length while retaining $\eta\sub{ext}=0.5$ by having $\Delta\sub{P}\approx g_0$.

As ever, there is a tradeoff to increasing birefringence in this manner, namely that the intra-cavity coupling between orthogonal polarisation eigenmodes will result in the emission of a photon with a time-dependent polarisation state.  This polarisation impurity is shown in the right hand figure of \cref{fig:TwoLevelAtomExtractionVsLength}a as the overall fraction of the cavity emission into the $\ket{\plus}$ mode.  The emitted time-dependent polarisation states are discussed in \cite{barrett19}, where it is shown that they can be modelled by using a Lindblad master equation approach.  In principle it is then possible to correct emitted photons back into a stationary polarisation mode by using, for example, a Pockels cell, if this was required for the overall application of the emitter-cavity system.

We emphasise that the effects we are describing are relevant to on-going experimental work.  As an example, we consider the system of Takahashi\etal{}~\cite{takahashi18} who demonstrated strong-coupling of an ion to an optical cavity.  With a $^{40}$Ca$^{+}$ ion coupled to a fibre-tip Fabry-Perot cavity, their system achieved coupling rates of $\{g_0,\kappa_0,\gamma\}/2\pi = \{12.3,4.1,11.5\}\si{\MHz}$ at $L_0=\SI{370}{\um}$.  Concurrently, multiple groups have demonstrated the controlled engineering of birefringence in such fibre-tip Fabry-Perot cavities through the ablation of elliptical mirror geometries~\cite{garcia18, cui18}.  \Cref{fig:TwoLevelAtomExtractionVsLength}b shows the achievable photon extraction efficiencies, once again using our model of a two-level emitter of circularly polarised light coupled to a linearly birefringent cavity, if birefringence was engineered into a cavity matching that of Takahashi\etal{}.  Significantly enhanced photon extraction can be obtained and the thick red contour denotes $\eta\sub{ext}=\SI{20.0}{\percent}$, the simulated performance of the physical system without birefringence.  With $\Delta\sub{P}/2\pi=\SI{18.4}{\MHz}$, the cavity could be lengthened by over \SI{50}{\percent} to $L\sub{cav}>\SI{550}{\um}$ without any loss of performance.  Alternatively, without altering the cavity length, the photon extraction efficiency is increased by a factor of $1.25$ to \SI{26.0}{\percent} at $\Delta\sub{P}/2\pi=\SI{19.3}{\MHz}$.

It is important to note that we do not consider other physical constraints to the achievable extraction rates.  These simulations simply underline that with the current state-of-the-art, the additional degree of freedom in cavity design afforded by birefringence can allow significantly different, and desirable, systems to be realised.

\bibliographystyle{apsrev4-1}
\bibliography{PushingPurcell_References}

\end{document}